\documentclass[amstex,aps,preprint,nofootinbib]{revtex4}%
\usepackage{graphicx}
\usepackage{color}
\usepackage{bm}
\usepackage{braket}
\usepackage{longtable}
\usepackage{amsmath}
\usepackage{amsfonts}
\usepackage{amssymb}

\begin{document}

\title{ Single particle nonlocality, geometric phases and time-dependent boundary conditions}
\author{A. Matzkin}
\affiliation{Laboratoire de Physique Th\'{e}orique et Mod\'{e}lisation (CNRS Unit\'{e}
	8089), Universit\'{e} de Cergy-Pontoise, 95302 Cergy-Pontoise cedex,
	France}

\begin{abstract}
	
We investigate the issue of single particle nonlocality in a quantum system subjected to time-dependent boundary conditions. We discuss earlier claims according to which the quantum state of a particle remaining localized at the center of an infinite well with moving walls would be specifically modified by the change in boundary conditions due to the wall's motion. We first prove that the evolution of an initially localized Gaussian state is not affected nonlocally by a linearly moving wall: as long as the quantum state has negligible amplitude near the wall, the boundary motion has no effect. This result is further extended to related confined time-dependent oscillators in which the boundary's motion is known to give rise to geometric phases: for a Gaussian state remaining localized far from the boundaries, the effect of the geometric phases is washed out and the particle dynamics shows no traces of a nonlocal influence that would be  induced by the moving boundaries.

\end{abstract}

\maketitle

\section{\textbf{Introduction}}

Quantum systems with time-dependent boundary conditions are delicate to
handle.\ Even the simplest system -- a particle in a box with infinitely high
but moving walls -- remains the object of ongoing investigations. From a
mathematical standpoint, a consistent and rigorous framework hinges on
unifying the infinite number of Hilbert spaces (one for each time $t$), each
endowed with its own domain of self-adjointness
\cite{dias2011,martino2013,art2015}. Explicit solutions have been found in
specific cases, notably for the infinite well with linear expanding or
contracting walls \cite{doescher}, later generalized to a family of confined
time-dependent linear oscillators whose frequency is related to the wall's
motion \cite{makowski91}. However general methods, such as employing a
covariant time derivative \cite{pershogin91} in order to track the change in
the boundary conditions or implementing a quantum canonical transformation
\cite{mosta1999} so as to map the time-dependent boundary conditions problem
to a fixed boundary problem with another Hamiltonian can at best give
perturbartive results. Explicit solutions call for numerical methods
\cite{glasser2009,fojon2010} but these are not well suited to investigate
fundamental effects, in particular when controversial effects are discussed.

This work precisely deals with a controversial effect, namely the existence of
possible nonlocal effects induced by time-dependent boundary conditions on a
quantum state remaining localized far from the boundaries. From a general standpoint, it
is known that systems with a cyclic evolution may display geometric phases, a
global property often said to be \textquotedblleft nonlocal\textquotedblright%
\ or \textquotedblleft holistic\textquotedblright. However it was initially
suggested by Greenberger \cite{greenberger}, and subsequently mentioned by
several authors, eg
\cite{makowski1992,zou2000,yao2001,wang2008,mousavi2012,mousavi2014}, that
time-dependent boundary conditions could give rise to a genuine form of
nonlocality: a particle at rest and localized in the center of the box,
remaining far from the moving walls, would be physically displaced by the
changing boundary conditions induced by the walls motion. This claim was never
shown rigorously to be exact (some arguments were given to support the idea of
nonlocality, sometimes in a hand waving fashion), but to the best of our
knowledge this claim was not shown to be incorrect either.

In this work we show that the moving walls have no effect on the dynamics of a
quantum state placed far from the wall. More precisely we prove that the
dynamics of a particle with an initial Gaussian wavefunction (the most widely
investigated case) does not depend on the boundary conditions as long as the
wavefunction remains negligible at the boundaries. This will first shown to be
the case in the infinite well with linearly moving walls, and we will then
extend these results to a family of related systems in which the moving
boundaries can give rise to geometric phases. The ingredients employed,
combining a time-dependent unitary transformation with a property of the
Jacobi theta functions, will be described in Sec. 2. Sec. 3 will deal with the
infinite potential well with linearly expanding walls, including the periodic
case with instantaneous reversal. Sec. 4 will investigate confined
time-dependent oscillators with a specific relation between the oscillator
frequency and the position of the confining walls; contrary to the infinite
potential well, such systems admit cyclic states displaying geometric phases
that will be seen to be induced by the wall's motion. We will nevertheless
show that the effect of the geometric phases is washed out when the initial
quantum state is localized inside the well. The results obtained will be
discussed in Sec. 5.

\section{\textbf{Quantum canonical transformation}}

\subsection{Hamiltonian and boundary conditions}

The Hamiltonian for a particle of mass $m$ placed in a potential $v(x,t)$
inside a confined\ well of width $L(t)$ with moving boundaries is given by%
\begin{align}
H  &  =\frac{P^{2}}{2m}+V\label{ham}\\
V(x,t)  &  =\left\{
\begin{array}
[c]{l}%
v(x,t)\text{ \ for}\ \ -\frac{L(t)}{2}\leq x\leq\frac{L(t)}{2}\\
+\infty\text{ \ otherwise}%
\end{array}
\right.  . \label{vdef}%
\end{align}
The solutions of the Schr\"{o}dinger equation
\begin{equation}
i\hbar\partial_{t}\psi(x,t)=H\psi(x,t)
\end{equation}
must obey the boundary conditions $\psi(\pm L(t)/2)=0$ (if the well is
embedded in a larger system more general boundary conditions can be considered
\cite{facchi}). The even and odd instantaneous eigenstates of $H$,
\begin{equation}
\phi_{n}(x,t)=\sqrt{2/L(t)}\cos\left[  \left(  2n+1\right)  \pi x/L(t)\right]
\label{even}%
\end{equation}
and%
\begin{equation}
\varphi_{n}(x,t)=\sqrt{2/L(t)}\sin\left[  \left(  2n\right)  \pi
x/L(t)\right]  \label{odd}%
\end{equation}
verify $H\ket{\phi_{n}}=E_{n}(t)\ket{\phi_{n}}$ and $H\ket{\varphi_{n}}=E_{n}%
(t)\ket{\varphi_{n}}.$ The instantaneous eigenvalues are $E_{n}(t)=\left(
2n+1\right)  ^{2}\hbar^{2}\pi^{2}/2mL^{2}(t)$ (with $n$ a positive integer)
and $E_{n}(t)=\left(  2n\right)  ^{2}\hbar^{2}\pi^{2}/2mL^{2}(t)$ (with $n$ a
strictly positive integer) for the even and odd instantaneous eigenstates
resp. However the $\phi_{n}$ or $\varphi_{n}$ are \emph{not} solutions of the
Schr\"{o}dinger equation. Indeed, due to the time-varying boundary conditions,
the problem is ill defined, eg the time derivative $\partial_{t}\psi(x,t)$
involves the difference of two vectors with different boundary conditions
belonging to different Hilbert spaces \cite{martino2013}. Hence neither the
difference $\psi(x,t^{\prime})-\psi(x,t)$ nor inner products taken at
different times $\left\langle \psi(t^{\prime})\right\vert \left.
\psi(t)\right\rangle $ are defined.

In the following we will restrict our discussion to symmetric boundary
conditions as specified by Eq (\ref{vdef}) and to initial states of even
parity in $x$ (in practice, states initially located at the center of the
box), so that only the even states $\phi_{n}(x,t)$ will come into play. The
reason for this choice is that the derivations are technically simpler and the
discussion more transparent. The extension to initial states with no definite
parity and to non-symmetric boundary conditions is given in the Appendix.

\subsection{Unitary transformation}

To tackle this problem the most straightforward approach is to map the
Hamiltonian $H$ of the time-dependent boundary conditions to a new Hamiltonian
$\tilde{H}$ of a fixed boundary problem. This is done by employing a
time-dependent unitary transformation implementing a \textquotedblleft
canonical\textquotedblright\ change of variables \cite{mosta1999}. Let
\begin{equation}
\mathcal{M}(t)=\exp\left(  \frac{i\xi(t)}{2\hbar}\left(  XP+PX\right)
\right)  \label{udef}%
\end{equation}
be a unitary operator with a time-dependent real function $\xi(t)$ defining
the canonical transformation \cite{mosta1999}%
\begin{align}
\ket{ \tilde{\psi}}  &  =\mathcal{M}(t)\left\vert \psi\right\rangle
\label{psitrt}\\
\tilde{H}(t)  &  =\mathcal{M}(t)H(t)\mathcal{M}^{\mathcal{\dagger}}%
(t)+i\hbar\mathcal{M}(t)\partial_{t}\mathcal{M}^{\mathcal{\dagger}%
}(t)\label{tfh}\\
\tilde{A}  &  =\mathcal{M}(t)A\mathcal{M}^{\mathcal{\dagger}}(t)
\end{align}
the latter holding for time-independent observables $A$ such as $X$ or $P$.
Note that $\mathcal{M}(t)$ represents a dilation, ie any arbitrary function
$f(x)$ \ transforms as $\mathcal{M}(t)f(x)=e^{\xi(t)/2}f(e^{\xi(t)}x)$. It is
therefore natural to choose $\xi(t)$ so that $\exp(\xi(t))= L(t)/L_{0} $ where
$L_{0}\equiv L(t=0)$ so as to map the original problem to the initial interval
$\left[  -L_{0}/2,L_{0}/2\right]  $, with%
\begin{equation}
\psi\left(  x,t\right)  =\left\langle x\right\vert \mathcal{M}%
^{\mathcal{\dagger}}(t)\ket{ \tilde{\psi}}=\sqrt{\frac{L_{0}}{L(t)}}%
\tilde{\psi}\left(  \frac{L_{0}}{L(t)}x,t\right)  . \label{psitr}%
\end{equation}
$\ket{ \tilde{\psi}}$ is the solution of the fixed boundary Hamiltonian
(\ref{tfh}) whose explicit form is%
\begin{equation}
\tilde{H}(t)=\frac{\tilde{P}^{2}}{2m}+V(\tilde{X})-\frac{\partial_{t}%
L(t)}{2L(t)}\left(  XP+PX\right)  . \label{hamt}%
\end{equation}

Eq. (\ref{psitr}) suggests to work with solutions of $\tilde{H}(t)$. This is
particularly handy when a set of complete solutions $\ket{\tilde{\psi }_{n}}$%
\ obeying the canonically transformed Schr\"{o}dinger equation
\begin{equation}
i\hbar\partial_{t}\ket{ \tilde{\psi}_{n}} =\tilde{H}\ket{ \tilde{\psi}_{n}}
\label{ccse}%
\end{equation}
are known: the initial state $\left\vert \psi(t_{0})\right\rangle $ is mapped
to $\ket{ \tilde{\psi}(t_{0})},$ which is evolved by expansion over the basis
functions $\ket{ \tilde{\psi}_{n}} $ before being transformed by the inverse
unitary transformation.

\section{Evolution of a localized state in an infinite potential well
\label{sec-loc}}

\subsection{Moving walls at constant velocity: basis solutions}

Let us now consider the infinite potential well corresponding to $v(x,t)=0$ in
Eq. (\ref{vdef}). We will assume throughout that the walls move at constant
velocity $q$, so that the wall's motion follows
\begin{equation}
L(t)=L_{0}+qt. \label{ll}%
\end{equation}
$q>0$ ($q<0$) corresponds to linearly expanding (contracting) walls. The
linear motion (\ref{ll}) has been indeed the main case studied in the context
of nonlocality induced by boundary conditions, due to the existence of a
complete basis of exact solutions of the canonically transformed
Schr\"{o}dinger equation (\ref{ccse}).\ These solutions were originally
obtained by inspection \cite{doescher}, or later from a change of variables in
the Schr\"{o}dinger differential equation \cite{makowski91}. From these
solutions it is straightforward to guess the basis functions $\ket{
\tilde{\psi}_{n}} $ of Eq (\ref{ccse}) that are found to be given by%
\begin{equation}
\tilde{\psi}_{n}(x,t)=\sqrt{\frac{2}{L_{0}}}e^{\frac{imx^{2}L(t)\left[
\partial_{t}L(t)\right]  }{2\hbar L_{0}^{2}}-i\hbar\pi^{2}(2n+1)^{2}\int
_{0}^{t}L(t^{\prime})^{-2}\,dt^{\prime}/2m}\cos\left(  \pi(2n+1)x/L_{0}%
\right)  \label{eigent}%
\end{equation}
where $n=0,1,2...$ For the linear motion (\ref{ll}), the integral immediately
yields%
\begin{equation}
\int_{0}^{t}\frac{1}{L(t^{\prime})^{2}}\,dt^{\prime}=\frac{t}{L_{0}(L_{0}%
+qt)}.
\end{equation}
As mentioned above the $\tilde{\psi}_{n}(x,t)$ are not eigenfunctions of
$\tilde{H}$, but they can be employed as a fundamental set of solutions in
order to obtain the state $\ket{\tilde{\psi}(t)}$ evolved from an arbitrary
initial state $\ket{\tilde{\psi}(t=0)}$ expressed as%
\begin{equation}
\ket{\tilde{\psi}(t)}=\sum_{n}%
\braket{\tilde{\psi_n}(t=0)|\tilde{\psi}(t=0)}\ket{\tilde{\psi_n}(t)}.
\label{td}%
\end{equation}
The solution $\psi(x,t)$ of the original problem with moving boundaries is
recovered from $\tilde{\psi}(x,t)$ through Eq. (\ref{psitr}). In particular,
each solution $\tilde{\psi}_{n}(x,t)$ is mapped into%
\begin{equation}
\psi_{n}(x,t)=\sqrt{\frac{2}{L(t)}}e^{\frac{imx^{2}\left[  \partial
_{t}L(t)\right]  }{2\hbar L(t)}-i\hbar\pi^{2}(2n+1)^{2}\int_{0}^{t}%
L(t^{\prime})^{-2}\,dt^{\prime}/2m}\cos\left(  \pi(2n+1)x/L(t)\right)  .
\label{psifund}%
\end{equation}

\subsection{Gaussian Evolution}

\subsubsection{Initial Gaussian}

Assume the initial wavefunction is a Gaussian of width $d,$%
\begin{equation}
\left\langle x\right\vert \left.  G(t=0)\right\rangle \equiv G(x,0)=\frac
{(1-i)e^{-\frac{x^{2}}{4d^{2}}}}{2^{3/4}\pi^{1/4}\sqrt{-id}} \label{gau0}%
\end{equation}
with a maximum at the center of the box ($x=0$) and with negligible amplitude
at the box boundaries $x=\pm L_{0}/2$. We will consider in the Appendix the
more general case of an initial Gaussian with arbitrary initial average
position and momentum, given by Eq. (\ref{agau}). $\left\vert
G(t=0)\right\rangle $ is expanded over the basis states $\ket{
\tilde{\psi}_{n}(t=0)}$ as per Eq. (\ref{td}) where $g_{n}%
(q)=\braket{\tilde{\psi}_{n}(t=0)|G(t=0)}$ is readily obtained analytically
from%
\begin{align}
g_{n}(q)  &  =\int_{-\infty}^{+\infty}\psi_{n}^{\ast}(x,0)G(x,0)dx\label{gn1}%
\\
&  =\frac{(1-i)2^{3/4}\pi^{1/4}}{\sqrt{-idl_{0}}\sqrt{\frac{1}{d^{2}}%
+\frac{2imq}{\hbar l_{0}}}}\exp\left(  -\frac{\pi^{2}d^{2}\hbar(2n+1)^{2}%
}{l_{0}\left(  \hbar l_{0}+2id^{2}mq\right)  }\right)  \label{gn2}%
\end{align}
The fact that the solutions $\psi_{n}(x,t)$ stretch (in the expanding case) as
time increases has been taken as an indication that the initial Gaussian would
also stretch provided the expansion is done adiabatically so that the
expansion coefficients $g_{n}$ remain unaltered \cite{greenberger}. Hence the
physical state of the particle would be changed nonlocally by the expansion,
although no force is acting on it.

We show however that the evolution of the initial Gaussian can be solved
exactly in the linear expanding or retracting cases by using Eqs.
(\ref{psitr}) and (\ref{td}), displaying no dependence of the time-evolved
Gaussian on the walls motion. The periodic case, in which the walls motion
reverses and starts contracting at $T/2$ so that $L(T)=L_{0}$ follows by
connecting the solutions at $t=T/2$.

\subsubsection{Sum in terms of Theta functions}

Our approach to this problem involves the use of special functions, the Jacobi
Theta functions, and a well-known peculiar property of these functions (the
Transformation theorem \cite{bellman}). Let us introduce the Jacobi Theta
function, $\vartheta_{2}(z,\kappa)$, defined here as%
\begin{equation}
\vartheta_{2}(z,\kappa)=2\sum_{n=0}^{\infty}e^{i\pi\kappa\left(  n+1/2\right)
^{2}}\cos\left[  \left(  2n+1\right)  z\right]  \label{theta2}%
\end{equation}
with $\operatorname{Im}(\kappa)>0$. It can be verified that the time evolved
solution $\tilde{\psi}(x,t)=\sum_{n}g_{n}(q)\tilde{\psi}_{n}(x,t)$ can be
summed to yield a theta function $\vartheta_{2}$, and that further applying
Eq. (\ref{psitr}) gives the wavefunction evolved from $G(x,0)$ as
\begin{equation}
\psi(x,t)=\frac{(1-i)\left(  2\pi\right)  ^{1/4}e^{\frac{imx^{2}\partial
_{t}L(t)}{2hL(t)}}\vartheta_{2}\left(  z,\kappa\right)  }{\sqrt{-idL_{0}%
L(t)}\sqrt{\frac{1}{d^{2}}+\frac{2im}{hL_{0}}\partial_{t}L(t)_{t=0}}}
\label{solTH}%
\end{equation}
with%
\begin{equation}
z=\frac{\pi x}{L(t)};\enskip\kappa=\frac{4\pi\hbar d^{2}}{L_{0}\left(
2d^{2}m\partial_{t}L(t)_{t=0}-i\hbar L_{0}\right)  }-\frac{2\pi\hbar}{m}%
\int_{0}^{t}\frac{1}{L(t^{\prime})^{2}}dt^{\prime} \label{ez}%
\end{equation}
In general $\psi$ as well as $z$ and $\kappa$ depend on $q$, the velocity of
the walls motion. We will explicitly denote this functional dependence, ie
$z(q),\kappa(q).$ Note that the particular case $q=0$ corresponds to static
walls with fixed boundary conditions.

\subsubsection{Comparing the static and expanding walls cases\label{secex}}

In order to compare the time evolved wavefunction in the static and moving
problems, let us compute $\psi(x,t;q=0)/\psi(x,t;q)$ which after some simple
manipulations takes the form
\begin{equation}
\frac{\psi(x,t;q=0)}{\psi(x,t;q)}=e^{\frac{iz^{2}(0)}{\pi\kappa(0)}%
-\frac{iz^{2}(q)}{\pi\kappa(q)}}\left(  \frac{\kappa(0)}{\kappa(q)}\right)
^{1/2}\frac{\vartheta_{2}\left(  z(0),\kappa(0)\right)  }{\vartheta_{2}\left(
z(q),\kappa(q)\right)  }. \label{rap}%
\end{equation}
We now prove that for the physical values of the parameters corresponding to a
localized Gaussian, this expression is unity. The first step is to use the
Jacobi transformation \cite{bellman}
\begin{equation}
\vartheta_{2}\left(  z,\kappa\right)  =\frac{e^{-iz^{2}/\kappa\pi}}{\left(
-i\kappa\right)  ^{1/2}}\vartheta_{4}\left(  \frac{z}{\kappa},-\frac{1}%
{\kappa}\right)  \label{jt}%
\end{equation}
for both $\vartheta_{2}$ functions of Eq. (\ref{rap}). $\vartheta_{4}$ is the
Jacobi Theta function defined by
\begin{equation}
\vartheta_{4}\left(  z,\kappa\right)  =\sum_{n=-\infty}^{\infty}\left(
-1\right)  ^{n}e^{i\pi\kappa n^{2}}e^{2inz}. \label{th4def}%
\end{equation}
Eq. (\ref{rap}) then becomes%
\begin{equation}
\frac{\psi(x,t;q=0)}{\psi(x,t;q)}=\frac{\vartheta_{4}\left(  \frac
{z(0)}{\kappa(0)},-1/\kappa(0)\right)  }{\vartheta_{4}\left(  \frac
{z(q)}{\kappa(q)},-1/\kappa(q)\right)  }. \label{argui1}%
\end{equation}

We then note that
\begin{equation}
\operatorname{Im}\left[  -1/\kappa(q)\right]  =d^{2}m^{2}L(t)^{2}/\pi\left(
4d^{4}m^{2}+h^{2}t^{2}\right)  . %
\end{equation}
This is typically a very large quantity, 
$\operatorname{Im}-1/\kappa(q)\gg1$. This comes from the fact that the typical spatial extension
$\Delta x$ of a Gaussian at time $t$ is deduced from its variance $\left(
\Delta x\right)  ^{2}$. $\Delta x$ needs to be much less than the
spatial extension of the well $L(t)$ since by assumption the quantum state
remains localized at the center of the box, far from the box boundaries.  Recall indeed that for a Gaussian
$\left(  \Delta x\right)  ^{2}=d^{2}+(\hbar t)^{2}/(2dm)^{2}$, so for
expanding walls the condition $\left(  \Delta x\right)  (t)\ll L(t)$ can be
fulfilled even for large $t$ provided $q$ is sufficiently large.\ However,
since we are comparing here the evolution for an arbitrary value of $q$ with
the fixed walls case ($q=0$), the stricter condition for $q=0$ 
\begin{equation}
\left(  \Delta x\right)  (t)\ll L_{0}
 \label{pich}
\end{equation}
 is the one that needs to hold. This condition will only hold for a limited 
time interval, given that the initially localized quantum state will spread and
necessarily reach the walls. But then of course the question regarding nonlocal
effects of the boundaries motion becomes moot, since a local contact with an infinite wall (be it fixed or moving) reflects the wavefunction and modifies its dynamics. 
This is why the investigation concerning nonlocal effects is only relevant for times such that Eq. (\ref{pich})  holds, although it should be stressed that the time
evolved expression for $\psi(x,t)$ that we have derived, given by Eq. (\ref{solTH}) remains valid
for any $t$.

 Now from the definition of $\vartheta_{4}$ we have%
\begin{equation}
\vartheta_{4}\left(  \frac{z(q)}{\kappa(q)},-\frac{1}{\kappa(q)}\right)
=\sum_{n=-\infty}^{\infty}\left(  -1\right)  ^{n}e^{i\left(  \pi
n^{2}-2nz(q)\right)  \left[  \operatorname{Re}\left(  \frac{-1}{\kappa
(q)}\right)  \right]  }e^{-\left(  \pi n^{2}-2nz(q)\right)  \left[
\operatorname{Im}\left(  \frac{-1}{\kappa(q)}\right)  \right]  }. \label{eau}%
\end{equation}
The last term of Eq. (\ref{eau}) is negligible except for $n=0$, ie
$\exp\left(  \pi n^{2}-2nz(q)\right)  \left[  \operatorname{Im}\left(
\frac{1}{\kappa(q)}\right)  \right]  \simeq0+\delta_{n,0}$ because $z(q)$ is
real, with $\left\vert z(q)\right\vert \ll1/2$ (since the spatial wavefunction
is assumed to vanish outside the central part of the well), and
$\operatorname{Im}(1/\kappa)<0.$ Therefore Eq. (\ref{eau}) is reduced to the
single term $n=0$ yielding $\vartheta_{4}\left(  z(q)/\kappa(q),-1/\kappa
(q)\right)  \simeq1$. This holds for \emph{any} value of $q$ and in particular
for $q=0$ (fixed walls). Hence, according to Eq. (\ref{rap}), we have
\begin{equation}
\psi(x,t;q)=\psi(x,t;q=0) \label{mr}%
\end{equation}
meaning that the dynamics of the wavefunction initially localized at the
center of the box does not depend on the expanding motion of the walls at the
boundaries of the box.\ In particular the adiabatic condition does not play
any particular role, as Eq. (\ref{mr}) holds for any value of the wall
velocity $q$. While each individual state $\psi_{n}(x,t)$ does stretch out as
time increases, the sum (\ref{td}) for $\psi(x,t)$ ensures that the
interferences cancel the stretching for the localized state. From a physical
standpoint no motion is induced superluminally on a localized quantum state by
the walls expansion.

\subsubsection{Contracting and periodic walls motion\label{iwp-per}}

The same results hold for walls contracting linearly (with now $q<0$),
provided the wavefunction remains localized far from the walls throughout .
The evolution in the periodic case follows by considering successively an
expansion with $L(t)=L_{0}+qt$ up to $t=T/2$ followed by a contraction from
$t=T/2$ to $T$ with the walls positions determined from
\begin{equation}
L^{c}(t)=L_{0}+q(T-t), \label{sinv}%
\end{equation}
now with $q>0$.\ The analytic solutions (\ref{eigent}) and (\ref{psifund}) do
not verify the Schr\"{o}dinger equation during the reversal. Assuming the
walls motion is instantaneously reversed at $t=T/2,$ the continuity of the
wavefunction imposes to match the expanding and contracting solutions at that
time. Note in particular that an expanding basis state $\psi_{n}%
(x,T/2-\varepsilon)$, where $\varepsilon$ is small, does not evolve into the
\textquotedblleft reversed\textquotedblright\ state $\psi_{n}%
(x,T/2+\varepsilon)$ after the walls motion reversal. Indeed the basis
solutions of the Schr\"{o}dinger equation with the contracting boundary
conditions given by Eq. (\ref{sinv}) are%
\begin{equation}
\psi_{n}^{c}(x,t)=\sqrt{\frac{2}{L_{0}+q(T-t)}}e^{\left(  -\frac{i\pi^{2}%
\hbar(2n+1)^{2}(2t-T)}{2m(2L_{0}+qT)(L_{0}+q\left(  T-t\right)  )}%
-\frac{imqx^{2}}{2\hbar(L_{0}+q(T-t))}\right)  }\cos\left(  \frac{\pi
(2n+1)x}{L_{0}+q(T-t)}\right)  , \label{cont}%
\end{equation}
and obviously $\psi_{n}(x,T/2)\neq\psi_{n}^{c}(x,T/2).$ We have instead a
diffusion process, in which a given basis function $\psi_{n}$ of the expanding
boundary condition is scattered into several outgoing channels $\psi_{j}^{c}$
of the contracting boundary case. This holds for any nonvanishing value of
$q$; to first order, we have%
\begin{equation}
\frac{\psi_{n}(x,T/2)}{\psi_{n}^{c}(x,T/2)}=1+iq\frac{mx^{2}}{\hbar L_{0}%
}+o(q^{2}),
\end{equation}
so that even in the adiabatic limit the expanding basis wavefunction cannot be
matched to a contracting one, as implicitly assumed in Ref. \cite{greenberger}.

In order to obtain evolved localized Gaussian in the periodic case, we can
proceed as follows. From Eq. (\ref{mr}) (taken for $q\rightarrow\infty$), we
know that $\psi(x,T/2;q)$ is a freely evolved Gaussian. We can thus repeat the
same steps leading to (\ref{solTH}), but starting from the time evolved
Gaussian%
\begin{equation}
G(x,T/2)=\frac{(1-i)e^{\frac{imx^{2}}{2\left(  \hbar T/2-2id^{2}m\right)  }}%
}{\left(  2\pi\right)  ^{1/4}\sqrt{d\left(  \frac{\hbar T/2}{d^{2}%
m}-2i\right)  }}%
\end{equation}
instead of Eq. (\ref{gau0}). $G(x,T/2)$ is the expanded over the contracting
basis functions $\tilde{\psi}_{n}^{c}(x,t)$ [cf Eq. (\ref{cont})], the
expansion coefficients $g_{n}^{c}(q)$ replacing the former $g_{n}(q)$
introduced above in Eq. (\ref{gn1}). The result is%
\begin{equation}
g_{n}^{c}(q)=\frac{(1-i)2^{3/4}\pi^{1/4}\exp\left(  -\frac{\hbar(2\pi
n+\pi)^{2}\left(  4d^{2}m+i\hbar T\right)  }{2m(2L_{0}+qT)\left(  \hbar
(L_{0}+qT)-2id^{2}mq\right)  }\right)  }{\sqrt{2L_{0}+qT}\sqrt{\frac{\hbar
T}{dm}-4id}\sqrt{\frac{m\left(  \hbar(L_{0}+qT)-2id^{2}mq\right)  }%
{h(2L_{0}+qT)\left(  4d^{2}m+i\hbar T\right)  }}}.
\end{equation}
The final step, as above, is to write the formal infinite sum in terms of the
Theta function $\vartheta_{2}$. At $t=T$, when the walls have recovered their
initial position $L(T)=L_{0}$, the time evolved Gaussian is given by
\begin{equation}
\psi^{c}(x,T;q)=\frac{(1-i)\left(  2\pi\right)  ^{1/4}e^{-\frac{imqx^{2}%
}{2\hbar L_{0}}}\vartheta_{2}\left(  \frac{\pi x}{L_{0}},\kappa^{c}(q)\right)
}{\sqrt{L_{0}\left(  \frac{T}{d}-\frac{4idm}{\hbar}\right)  }\sqrt
{\frac{-2id^{2}mq+\hbar L_{0}+\hbar qT}{4d^{2}m+i\hbar T}}} \label{psitff}%
\end{equation}
where $\kappa^{c}(q)$ (at time $t=T$) is given by%
\begin{equation}
\kappa^{c}(q)=-\frac{2\pi\hbar\left(  \hbar T-2id^{2}m\right)  }{L_{0}m\left(
\hbar(L_{0}+qT)-2id^{2}mq\right)  }.
\end{equation}
We then use the same method that led us from Eq. (\ref{rap}) to Eq. (\ref{mr})
based on the Jacobi transformation theorem to show that $\psi^{c}%
(x,T;q)=\psi^{c}(x,T;q=0),$ that is the walls motion after a full cycle has no
consequence on the dynamics of a localized quantum state of the particle.

\section{Effect of geometric phases on a localized state evolution}

\subsection{Geometric phases and nonlocality}

For the infinite potential well with moving boundaries, the fact that the
basis functions are not cyclic states even in the case of periodic motion of
the walls (as seen in Sec. \ref{iwp-per}) precludes the existence of a cyclic
non-adiabatic geometric phase \cite{aharonov-anandan}.\ However geometric
phases \cite{book-GP} could be relevant to the issue of nonlocality. Indeed, a
geometric phase is a global quantity, affecting the quantum state globally
even if the effect causing the geometric phase lies in a localized space-time
region (we will see an explicit example below). Some authors even ascribe to
geometric phases nonlocal properties \cite{anandan} including in the context
of time-dependent boundary conditions \cite{wang2008}.

For these reasons it is relevant to see if the results obtained in Sec.
\ref{sec-loc} for the infinite potential well could be affected in systems
admitting geometric phases. It turns out that there are systems, a family of
time-dependent linear oscillators (TDLO) confined by infinitely high moving
walls, whose solutions are closely related to the ones of the infinite well
with time-dependent boundary conditions, that admit cyclic states that pick up
geometric phases. We will see by using a simple scaling property that the
geometric phase in this system is caused by the walls motion, but that
nevertheless the geometric phases have no consequence on the dynamics of a
localized quantum state.

\subsection{Confined time-dependent oscillators: geometric phase and basis
states}

\subsubsection{Confined TDLO}

Let us start again from the Hamiltonian (\ref{ham}) but now take $v(x,t)$ of
Eq. (\ref{vdef}) to be given by
\begin{equation}
v(x,t)=-\frac{m}{2}\frac{\partial_{t}^{2}L(t)}{L(t)}x^{2}.\label{vmod}%
\end{equation}
This is a TDLO confined in the interval$\ \ -L(t)/2\leq x\leq L(t)/2,$ where
as above $L(t)$ represents the size of the box between infinitely high and
moving walls. This TDLO is special in that the frequency $\Omega
^{2}(t)=-\partial_{t}^{2}L(t)/L(t)$ depends on the walls motion\footnote{Note
that when $L(t)$ is linear in $t,\partial_{t}^{2}L(t)$ vanishes and the
potential (\ref{vmod}) becomes that of the infinite well.\ Hence in general Eq.
(\ref{psifund}) represents the solution of a confined TDLO, only the special
case for which $\partial_{t}^{2}L(t)=0$ corresponds to the infinite well with
moving walls.}. It is then known \cite{makowski1992,mousavi2012}, as can be
checked directly by inspection, that the functions $\tilde{\psi}_{n}(x,t)$ and
$\psi_{n}(x,t)$ defined respectively by Eqs. (\ref{eigent}) and (\ref{psifund}%
) still obey the Schr\"{o}dinger equations $i\hbar\partial_{t}\tilde{\psi
}=\tilde{H}\tilde{\psi}$ and $i\hbar\partial_{t}\psi=H\psi$ where the
potential between the walls is now given by Eq. (\ref{vmod}).

\subsubsection{Cyclic Evolution}

Assume a confined TDLO with a real and periodic function $L(t)$ with period
$T$ is initially in a state $\psi_{n}(x,t=0)$, given by Eq. (\ref{psifund}).
After a full cyclic evolution $\psi_{n}(x,T)$ returns to the initial $\psi
_{n}(x,0)$ but acquires a total phase $\mu_{n},$ ie%
\begin{equation}
\psi_{n}(x,T)=e^{-i\mu_{n}}\psi_{n}(x,0).\label{mufaz}%
\end{equation}
Following Aharonov and Anandan \cite{aharonov-anandan}, $\mu_{n}$ can be
parsed into a \textquotedblleft dynamical\textquotedblright\ part $\delta_{n}$
encapsulating the usual phase increment by the instantaneous expectation value
of the Hamiltonian and a \textquotedblleft geometric\textquotedblright\ part
$\gamma_{n}$ reflecting the curve traced during the evolution in the
projective Hilbert space (defined as the space comprising the rays, that is
the states giving rise to the same density matrix
\cite{aharonov-anandan,book-GP}). $\mu_{n}$ is directly obtained from Eq.
(\ref{mufaz}) [with Eq. (\ref{psifund})] and is seen to be proportional to
$\int_{0}^{T}L(t^{\prime})^{-2}\,dt^{\prime}.$ The dynamical phase%
\begin{equation}
\delta_{n}=-\hbar^{-1}\int_{0}^{T}\left\langle \psi_{n}(t^{\prime})\right\vert
H\left\vert \psi_{n}(t^{\prime})\right\rangle dt^{\prime}%
\end{equation}
is computed through a tedious but straightforward calculation. The
nonadiabatic geometric phase $\gamma_{n}$ is then obtained as
\begin{equation}
\gamma_{n}=\mu_{n}-\delta_{n}=\frac{m}{24}\left(  1-\frac{6}{(2\pi n+\pi)^{2}%
}\right)  \int_{0}^{T}\left(  \partial_{t}L(t)\right)  ^{2}-L(t)\partial
_{t}^{2}L(t)dt.\label{faz}%
\end{equation}
Note that $\gamma_{n}$ is nonzero for nontrivial choices of $L(t)$.

\subsubsection{Scaling}

In order to get a handle on the physical origin of the geometric phase on a
state $\psi_{n}$, we use the following scaling property. By rescaling $L(t),$
$\bar{L}(t)=kL(t)$, with $k>1$ one changes the walls position while leaving
the dynamics invariant: put $\bar{L}(t)=kL(t)$, with $k>1.$ Then the frequency
$\Omega^{2}(t)=-\partial_{t}^{2}L(t)/L(t)$ and therefore the Hamiltonian are
not modified, by virtue of Eq. (\ref{vmod}). However as is apparent from Eq.
(\ref{faz}) the geometric phase scales as $\bar{\gamma}_{n}=k^{2}\gamma_{n}$.
Hence increasing the walls motion by a factor $k$ induces a change in the
geometric phase on the basis states $\bar{\psi}_{n}(x,T)$ that can be detected
at any point $x$ inside the confined oscillator.

\begin{figure}[tb]
\includegraphics[height=7cm]{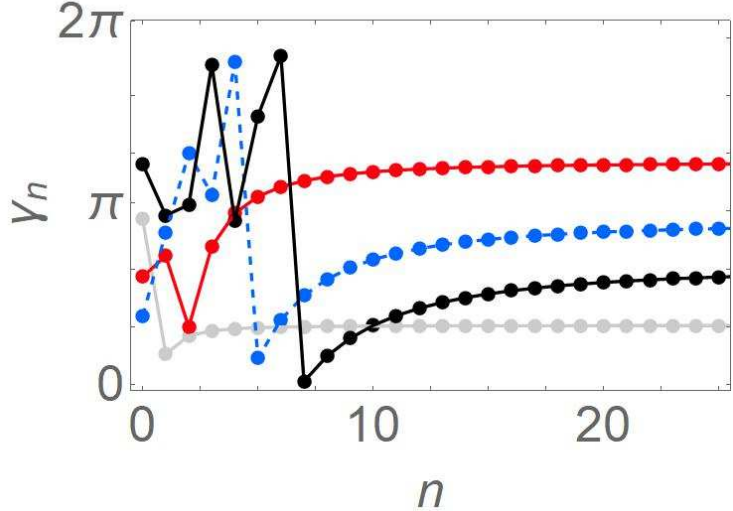}\caption{The geometric phase $\gamma
_{n}$ [cf Eq. (\ref{faz})] is given as a function of $n$ for the lowest basis
states $\psi_{n}$ ($\gamma_{n}$ is plotted $\operatorname{mod}2\pi$). The
system is a time-dependent oscillator with the frequency $\Omega(x,t)$ and the
walls motion $L(t)$ given by Eqs. (\ref{freqex}) and (\ref{ltex}) resp. The
black, blue dotted and red curves correspond to $\bar{L}_{0}=1000,800$, and
400 resp. while $\bar{L}_{0}=100$ is shown in light gray. Put differently, the
curves correspond to $L_{0}=100$ and the scaling parameter $k=4$, $8$ and 10
($q=0.1,\omega=1$, units with $\hbar,m,e=1$ are used). }%
\label{figGP}%
\end{figure}

This is illustrated in Fig.\ \ref{figGP} featuring a TDLO with%
\begin{equation}
L(t)=L_{0}\left(  \frac{\left(  1+q\right)  }{\left(  1+q\cos\omega t\right)
}\right)  ^{1/2} \label{ltex}%
\end{equation}
and the frequency in Eq. (\ref{vmod}) is given by%
\begin{equation}
\Omega^{2}(t)=-\partial_{t}^{2}L(t)/L(t)=\frac{q\omega^{2}(q(\cos(2\omega
t)-5)-4\cos(\omega t))}{8(q\cos(\omega t)+1)^{2}}. \label{freqex}%
\end{equation}
This particular choice of $L(t)$ for the boundary motions has been previously
investigated and is known in the infinite potential well case to lead to
chaotic or regular behavior as $L_{0},q$ and $\omega$ are varied
\cite{glasser2009}.\ Here instead we are looking at the confined TDLO, ie with
the potential given by Eq. (\ref{vmod}): Fig.\ \ref{figGP} shows the geometric
phase $\gamma_{n},$ computed from Eq. (\ref{faz}) for the first basis states
$\psi_{n}$ and for different values of $\bar{L}_{0}=kL_{0}$ thus illustrating
the dependence of $\gamma_{n}$ on the walls motion.

\subsection{Confined time-dependent oscillators: Localized state}

\subsubsection{Evolution of Gaussian state}

The time-dependent boundary conditions induce geometric phases on the basis
states $\psi_{n}$. Although a Gaussian state $G(x,0)$ initially given by Eq.
(\ref{gau0}) can be expanded at any time in terms of these basis states
$\psi_{n}$ this does not imply of course that the evolved wavefunction
$\psi(x,t)$ will also pick up a phase after a full cycle.

Actually, since Eqs. (\ref{eigent}) and (\ref{psifund}) still hold for the
confined TDLO with moving walls, we can again write the time-evolved solution
$\psi(x,T)$, here after a period $T$ in terms of a Theta function. Formally
$\psi(x,t)$ is again given by Eq. (\ref{solTH}), the only difference relative
to the infinite potential well of Sec. \ref{sec-loc} being that $L(t)$ is a
periodic function and not linear in $t$. To assess the relevance of geometric
phases, we rescale the walls motion while leaving the Hamiltonian invariant as
explained above by putting $\bar{L}(t)=kL(t)$. We have seen that this
rescaling modifies the geometric phases. Hence by comparing the rescaled
wavefunction $\bar{\psi}(x,T)$ with the original solution $\psi(x,T),$ evolved
in both cases from the same initial state $G(x,0),$ we can infer whether the
geometric phases modify the quantum state evolution.

Writing $\bar{\psi}(x,T)/\psi(x,T)$ in terms of $\vartheta_{4}$ functions as
per Eq. (\ref{solTH}), and noting that $\bar{z}=z/k$, $\bar{\kappa}%
=\kappa/k^{2}$ and $L(T)=L_{0},$ we apply the Jacobi transformation (\ref{jt})
to find given by Eq. (\ref{gau0}). From Eq. (\ref{ez}) we see that $\bar
{z}=z/k$ and $\bar{\kappa}=\kappa/k^{2}$ so that by using Eq. (\ref{solTH})
and the Jacobi transformation (\ref{jt}) we are led to%
\begin{equation}
\frac{\bar{\psi}(x,T)}{\psi(x,T)}=\frac{\vartheta_{4}\left(  \frac{1}{k}%
\frac{z}{\kappa},-k^{2}/\kappa\right)  }{\vartheta_{4}\left(  \frac{z}{\kappa
},-1/\kappa\right)  }.\label{gg}%
\end{equation}
The equality on the right handside holds only provided the conditions given
above between Eqs. (\ref{argui1}) and (\ref{mr}) hold (recall we have $k>1$).
Under these circumstances we see, by following exactly the reasoning given
above that both $\vartheta_{4}\left(  \frac{1}{k}\frac{z}{\kappa}%
,-k^{2}/\kappa\right)  \simeq1$ and $\vartheta_{4}\left(  \frac{z}{\kappa
},-1/\kappa\right)  \simeq1.$ 

Eq. (\ref{gg}) proves that while rescaling the walls motion changes the
geometric phase of the basis functions according to $\bar{\gamma}_{n}%
=k^{2}\gamma_{n}$, no such change takes place when the initial state is the
Gaussian $G(x,0)$ localized at the center of the confined time-dependent
potential.\ The geometric phases picked up by each basis state over which
$G(x,0)$ is expanded vanish by interference. Recall that an arbitrary initial
Gaussian placed in a periodic (unconfined) potential is not cyclic unless
specific conditions are verified \cite{child98}. Eq. (\ref{gg}) does not
depend on whether these conditions are met and suggests that the wavefunction
in the time-dependent boundary problem follows the same evolution as the one
of the unconfined problem with the time-dependent potential: $\psi(x,T)$ can
thus pick up a nonadiabatic cyclic geometric phase if the evolution in the
unconfined potential leads to such a geometric phase, but there will be no
additional effect due to the time-dependent boundaries.\ 

\subsubsection{Approximate solution for an unconfined time-dependent
oscillator}

Note that as a byproduct of the present treatment, we have obtained an
interesting closed form expression for the evolution of an initial Gaussian in
an unconfined time-dependent linear oscillator potential for which there is a
function $L(t)$ such that the frequency can be put under the form $\Omega
^{2}(t)=-\partial_{t}^{2}L(t)/L(t)$. Indeed, Eq. (\ref{solTH}) along with the
Jacobi transformation (\ref{jt}) and $\vartheta_{4}\left(  \frac{z}{\kappa
},-1/\kappa\right)  \simeq1$ give the evolved Gaussian $\psi(x,t)$ as%
\begin{equation}
\psi(x,t)=\frac{(1-i)\left(  2\pi\right)  ^{1/4}e^{\frac{imx^{2}\partial
_{t}L(t)}{2hL(t)}}}{\sqrt{-idL_{0}L(t)}\sqrt{\frac{1}{d^{2}}+\frac{2im}%
{hL_{0}}\partial_{t}L(t)_{t=0}}}\frac{e^{ix^{2}/\left[  4\hbar L(t)^{2}\left(
\frac{\tau(t)}{2m}+\frac{d^{2}}{-2d^{2}mL_{0}\partial_{t}L(t)_{t=0}+i\hbar
L_{0}^{2}}\right)  \right]  }}{\sqrt{\frac{4\pi d^{2}\hbar}{L_{0}\left(  \hbar
L_{0}+2id^{2}m\partial_{t}L(t)_{t=0}\right)  }+\frac{2i\pi\hbar\tau(t)}{m}}%
},\label{approx}%
\end{equation}
where $\tau(t)\equiv\int_{0}^{t}L^{-2}(t^{\prime})dt^{\prime}$. Contrary to
the standard approaches for solving Gaussian problems in TDLOs, that involve
nonlinear equations calling for numerical integration
\cite{child98,matzkin2012}, Eq. (\ref{approx}) can be often obtained
explicitly analytically, depending on whether the closed form integral of
$\tau(t)$ is known (of course the range of application of Eq. (\ref{approx})
is very limited compared to standard methods).

\subsubsection{Example}

\begin{figure}[tb]
\includegraphics[height=5cm]{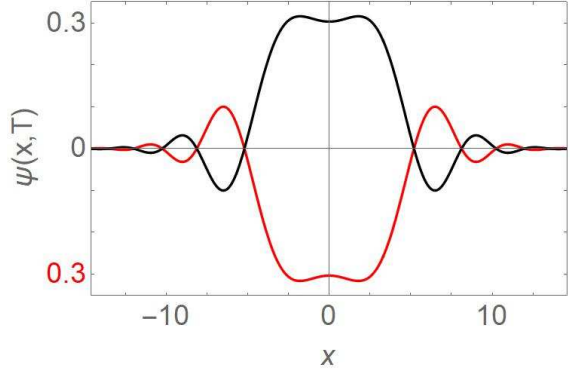}\caption{The state $\psi(x,t=T)$ evolved
from an initial Gaussian [Eq. (\ref{gau0}), here with $d=1$] in a confined
oscillator with the time-dependent frequency given by Eq. (\ref{freqex}) and
the walls moving according to Eq. (\ref{ltex}) with $L_{0}=100$ is compared to
the state at $t=T$ evolved from the same initial Gaussian in the unconfined
case (same Hamiltonian but without confining walls). The black line shows
$\operatorname{Re}\psi(x,t)$ at $t=T$ (after one full cycle); the red line
(upside-down) shows the evolution for the \emph{unconfined} TDLO. }%
\label{figGP2}%
\end{figure}

Let us look at the localized state evolution for the TDLO whose geometric
phases in the basis states were shown in Fig.\ \ref{figGP}. We start with an
initial Gaussian state and let it evolve up to $t=T$ for the TDLO confined by
infinitely high moving walls on the one hand, and for the same but unconfined
TDLO on the other. Fig.\ \ref{figGP2} shows the real part of the evolved
wavefunction in both cases. The curves are identical, illustrating that the
walls motion has no influence on the evolution of a localized state. Note that
the wavefunction for the unconfined TDLO has been computed by employing an
independent and totally different method, based on Gaussian propagation
through the solutions of Ermakov systems (see Ref. \cite{A2015JPA} for details).

\section{\textbf{Conclusion}}

To sum up we have shown that contrary to earlier claims, time-dependent
boundary conditions do not induce an effective or explicit form of
nonlocality, as happens eg for Bell correlations.\ This was seen to be the
case for the paradigmatic particle in a box with moving walls and also holds
for systems in which the moving boundaries induce geometric phases. Although a
moving wall changes the boundary conditions, this change modifies the entire
quantum state of the system (instantaneously in the non-relativistic
framework) only if the state has a non-negligible amplitude in the boundary
region. This is clearly not the case for a localized state placed far from the
moving walls.

\appendix



\section{\textbf{{Time evolved state in terms of theta functions}}}

We first consider the case of an initial state given by the Gaussian%
\begin{equation}
\left\langle x\right\vert \left.  \mathcal{G}(t=0)\right\rangle \equiv
\mathcal{G}(x,0)=\frac{e^{-\frac{(x-x_{0})^{2}}{4d^{2}}+\frac{ip_{0}x}{h}}%
}{\left(  2\pi\right)  ^{1/4}\sqrt{d}}. \label{agau}%
\end{equation}
Contrary to the initial Gaussian given by Eq. (\ref{gau0}), $\mathcal{G}(x,0)$
has its maximum at $x_{0}$ anywhere inside the box (but sufficiently far from
the box boundaries, since by assumption the initial state has negligible
amplitude at the boundaries), and a mean momentum $p_{0}$. In addition to the
even basis functions (\ref{eigent}) and (\ref{psifund}) derived from the even
instantaneous eigenstates $\phi_{n}(x,t)$ given by Eq. (\ref{even}), we will
also need odd basis functions derived in the same way from the odd eigenstates
$\varphi_{n}(x,t)$ given by Eq. (\ref{odd}); for example the odd counterpart
to $\psi_{n}(x,t)$ defined by Eq. (\ref{psifund}) is%
\begin{equation}
\zeta_{n}(x,t)=\sqrt{\frac{2}{L(t)}}e^{\frac{imx^{2}\left[  \partial
_{t}L(t)\right]  }{2\hbar L(t)}-i\hbar\pi^{2}(2n)^{2}\int_{0}^{t}L(t^{\prime
})^{-2}\,dt^{\prime}/2m}\sin\left(  \pi(2n)x/L(t)\right)  . \label{psifundodd}%
\end{equation}
The expansion coefficients $g_{n}$ of Eq. (\ref{gn1}) now become%
\begin{align}
h_{n}(q)  &  =\int_{-\infty}^{+\infty}\psi_{n}^{\ast}(x,0)\mathcal{G}(x,0)dx\\
&  =\frac{(1-i)\left(  -\pi/2\right)  ^{1/4}e^{-\frac{x_{0}^{2}}{4d^{2}}%
}\left(  e^{\frac{i\left[  2\pi d^{2}\hbar(2n+1)+\left(  2d^{2}p_{0}-i\hbar
x_{0}\right)  L_{0}\right]  ^{2}}{4d^{2}\hbar L_{0}\left(  2d^{2}m\partial
_{t}L(t)_{t=0}-i\hbar L_{0}\right)  }}+e^{\frac{i\left[  2\pi d^{2}%
\hbar(2n+1)+\left(  -2d^{2}p_{0}+i\hbar x_{0}\right)  L_{0}\right]  ^{2}%
}{4d^{2}\hbar L_{0}\left(  2d^{2}m\partial_{t}L(t)_{t=0}-i\hbar L_{0}\right)
}}\right)  }{\sqrt{dL_{0}}\sqrt{\frac{1}{d^{2}}+\frac{2im\partial
_{t}L(t)_{t=0}}{\hbar L_{0}}}}%
\end{align}
for the even basis functions and
\begin{align}
j_{n}(q)  &  =\int_{-\infty}^{+\infty}\zeta_{n}^{\ast}(x,0)\mathcal{G}%
(x,0)dx\label{intj}\\
&  =\frac{i\left(  2\pi\right)  ^{1/4}e^{-\frac{x_{0}^{2}}{4d^{2}}}%
e^{\frac{i\left[  2\pi\hbar d^{2}(2n)+\left(  -2d^{2}p_{0}+i\hbar
x_{0}\right)  L_{0}\right]  ^{2}}{4d^{2}\hbar L_{0}\left(  2d^{2}m\partial
_{t}L(t)_{t=0}-i\hbar L_{0}\right)  }}-e^{^{\frac{i\left[  2\pi\hbar
d^{2}(2n)+\left(  2d^{2}p_{0}-i\hbar x_{0}\right)  L_{0}\right]  ^{2}}%
{4d^{2}\hbar L_{0}\left(  2d^{2}m\partial_{t}L(t)_{t=0}-i\hbar L_{0}\right)
}}}}{\sqrt{dL_{0}}\sqrt{\frac{1}{d^{2}}+\frac{2im\partial_{t}L(t)_{t=0}}{\hbar
L_{0}}}} \label{intj2}%
\end{align}
for the odd basis functions.

It can be checked, after a tedious but straightforward calculation that the
time evolved state $\psi(x,t)=\sum_{n\geqslant0}h_{n}(q)\psi_{n}%
(x,t)+\sum_{n>0}j_{n}(q)\zeta_{n}(x,t)$ can be written in terms of 8 Jacobi
theta functions, half of them being theta functions of the second type
$\vartheta_{2}(z,\kappa)$ introduced above [Eq. (\ref{theta2})], the other
half (for the odd part of the sum) being functions $\vartheta_{3}(z,\kappa)$
defined by
\begin{equation}
\vartheta_{3}(z,\kappa)=\sum_{n=-\infty}^{\infty}e^{i\pi\kappa n^{2}}e^{2inz}.
\label{th3def}%
\end{equation}
Put%
\begin{align}
A  &  \equiv\exp\left(  -\frac{x_{0}^{2}}{4d^{2}}+\frac{i\left(  2d^{2}%
p_{0}-i\hbar x_{0}\right)  ^{2}L_{0}}{4d^{2}\hbar\left(  2d^{2}m\partial
_{t}L(t)_{t=0}-i\hbar L_{0}\right)  }+\frac{imx^{2}\partial_{t}L(t)_{t=0}%
}{2\hbar L_{0}}\right) \\
B  &  \equiv\sqrt{dL_{0}L(t)}\sqrt{\frac{1}{d^{2}}+\frac{2im\partial
_{t}L(t)_{t=0}}{\hbar L_{0}}}\\
C  &  \equiv\pi\frac{2d^{2}p_{0}-i\hbar x_{0}}{i\hbar L_{0}-2d^{2}%
m\partial_{t}L(t)_{t=0}}.
\end{align}
Note that $A$ and $C$ depend on $x_{0}$ and $p_{0}$. With $\kappa$ defined by
Eq. (\ref{ez}) above, we introduce the functions%
\begin{align}
\theta_{1}(x,t;q)  &  =(1-i)\left(  -\pi\right)  ^{1/4}A\vartheta_{2}%
(-\frac{\pi x}{L(t)}-C,\kappa)/B\label{t1}\\
\theta_{2}(x,t;q)  &  =(1-i)\left(  -\pi\right)  ^{1/4}A\vartheta_{2}%
(\frac{\pi x}{L(t)}-C,\kappa)/B\\
\theta_{3}(x,t;q)  &  =(1-i)\left(  -\pi\right)  ^{1/4}A\vartheta_{2}%
(-\frac{\pi x}{L(t)}+C,\kappa)/B\\
\theta_{4}(x,t;q)  &  =(1-i)\left(  -\pi\right)  ^{1/4}A\vartheta_{2}%
(\frac{\pi x}{L(t)}+C,\kappa)/B\label{t4}\\
\theta_{5}(x,t;q)  &  =\left(  \frac{\pi}{2}\right)  ^{1/4}A\vartheta
_{3}(-\frac{\pi x}{L(t)}-C,\kappa)/B\label{t5}\\
\theta_{6} (x,t;q)  &  =-\left(  \frac{\pi}{2}\right)  ^{1/4}A\vartheta
_{3}(\frac{\pi x}{L(t)}-C,\kappa)/B\\
\theta_{7}(x,t;q)  &  =-\left(  \frac{\pi}{2}\right)  ^{1/4}A\vartheta
_{3}(-\frac{\pi x}{L(t)}+C,\kappa)/B\\
\theta_{8}(x,t;q)  &  =\left(  \frac{\pi}{2}\right)  ^{1/4}A\vartheta
_{3}(\frac{\pi x}{L(t)}+C,\kappa)/B. \label{t8}%
\end{align}
Then the time-evolved state $\psi(x,t)$ analogous to the one obtained above
[Eq. (\ref{solTH})] but when the initial state is the general Gaussian given
by Eq. (\ref{agau}) is given in terms of the functions $\theta_{k}$ as%
\begin{equation}
\psi(x,t;q)=\frac{1}{2}\sum_{k=1}^{8}\theta_{k}(x,t;q). \label{psiGG}%
\end{equation}

\section{\textbf{{Moving walls at constant velocity}}}

Let us assess the effect of walls moving at constant velocity, discussed in
Sec. \ref{sec-loc}, on the wavefunction evolving from $\mathcal{G}(x,0)$. For
each of the even functions $\theta_{k}$ ($k=1,..,4)$ the transformation
(\ref{jt}) leads to the analog of Eq. (\ref{argui1}) in the form
\begin{equation}
\frac{\theta_{k}(q=0)}{\theta_{k}(q)}=\frac{\vartheta_{4}\left(  \frac
{z_{k}(0)}{\kappa(0)},-1/\kappa(0)\right)  }{\vartheta_{4}\left(  \frac
{z_{k}(q)}{\kappa(q)},-1/\kappa(q)\right)  },\qquad k=1,..,4
\end{equation}
where $z_{k}$ is the relevant argument of the theta function in the expression
of $\theta_{k}$ given by Eqs. (\ref{t1})-(\ref{t4}), that is $z_{k}=\pm
\frac{\pi x}{L(t)}\pm C$. \ As explained in Sec.\ \ref{secex} in the case of a
single theta function, this leads here, under the same assumptions, to
$\theta_{k}(x,t;q=0)=\theta_{k}(x,t;q)$, so that the walls motion does not
impinge on the evolution of each of these even functions $\theta_{k}$.

For the odd functions $\theta_{k}$ ($k=5,..,8)$ involving $\vartheta_{3},$ we
use instead of Eq. (\ref{jt}) the Jacobi transformation \cite{bellman}%
\begin{equation}
\vartheta_{3}\left(  z,\kappa\right)  =\frac{e^{-iz^{2}/\kappa\pi}}{\left(
-i\kappa\right)  ^{1/2}}\vartheta_{3}\left(  \frac{z}{\kappa},-\frac{1}%
{\kappa}\right)  .
\end{equation}
The result
\begin{equation}
\frac{\theta_{k}(q=0)}{\theta_{k}(q)}=\frac{\vartheta_{3}\left(  \frac
{z_{k}(0)}{\kappa(0)},-1/\kappa(0)\right)  }{\vartheta_{3}\left(  \frac
{z_{k}(q)}{\kappa(q)},-1/\kappa(q)\right)  }=1,\qquad k=5,..,8 \label{resodd}%
\end{equation}
is shown to hold by following the same arguments given in Sec.\ \ref{secex},
but by using the expansion (\ref{th3def}) instead of (\ref{th4def}). Thus Eq.
(\ref{mr}) above stating that $\psi(x,t;q)=\psi(x,t;q=0)$ also holds when the
initial state is the Gaussian (\ref{agau}) and $\psi(x,t;q)$ is given by Eq.
(\ref{psiGG}).

\section{\textbf{{A single moving wall}}}

In the main text we have considered the symmetric boundary conditions
specified by Eq (\ref{vdef}), as this gives a simpler treatment.\ However in
most of the works
\cite{greenberger,makowski1992,zou2000,yao2001,wang2008,mousavi2012,mousavi2014}
dealing with the subject of nonlocality induced by time-dependent boundary
conditions, the problem of an infinite well with a single moving wall was
considered. In that case, the Hamiltonian has the following boundary
conditions:
\begin{align}
H  &  =\frac{P^{2}}{2m}+V\\
V(x,t)  &  =\left\{
\begin{array}
[c]{l}%
0\text{ \ for}\ \ 0\leq x\leq L(t)\\
+\infty\text{ \ otherwise}%
\end{array}
\right.  .
\end{align}
The instantaneous eigenstates of $H$ are similar to the odd functions
$\varphi_{n}(x,t)$ introduced in Eq. (\ref{odd}) and the basis functions to
the $\zeta_{n}(x,t)$ of Eq. (\ref{psifundodd}); they are obtained by replacing
in these expressions $n$ by $n/2,$ yielding%
\begin{equation}
f_{n}(x,t)=\sqrt{2/L(t)}\sin\left[  n\pi x/L(t)\right]
\end{equation}
for the instantaneous eigenstates and%
\begin{equation}
F_{n}(x,t)=\sqrt{\frac{2}{L(t)}}e^{\frac{imx^{2}\left[  \partial
_{t}L(t)\right]  }{2\hbar L(t)}-i\hbar\pi^{2}n^{2}\int_{0}^{t}L(t^{\prime
})^{-2}\,dt^{\prime}/2m}\sin\left(  \pi nx/L(t)\right)
\end{equation}
for the basis functions. Therefore, provided we are willing to keep the
$-\infty$ bound in Eq. (\ref{intj}), a harmless approximation given the
assumptions concerning the initial Gaussian, we can transpose the results
obtained in the present Appendix [Eqs. (\ref{intj2}), (\ref{t5})-(\ref{t8})
and (\ref{resodd})] to the case of a single moving wall (note that relative to
these expressions, the arguments of $\vartheta_{3}$ are rescaled as
$z\rightarrow z/2$ and $\kappa\rightarrow\kappa/4$). Hence the conclusion
concerning the non-relevance of the wall's motion relative to the evolution of
a state compactly localized inside the box also holds in this case.


\vspace{2cm}

\begin{thebibliography}{99}                                                                                               %


\bibitem {dias2011}N. C. Dias, A. Posilicano and J. N. Prata 2011, Commun.
Pure Appl. Anal. 10 1687.

\bibitem {martino2013}S. Di Martino, F. Anza, P. Facchi, A. Kossakowski, G.
Marmo, A. Messina, B. Militello and S. Pascazio 2013, J. Phys. A 46 365301.

\bibitem {art2015}E. Knobloch and R. Krechetnikov 2015, Acta Appl. Math. 137 123.

\bibitem {doescher}S. W. Doescher and H. H. Rice 1969, Am. J. Phys. 37 1246

\bibitem {makowski91}A.J. Makowski and S.T. Dembinski 1991, Phys. Lett. A 154 217

\bibitem {pershogin91}P. Pereshogin and P. Pronin 1991, Phys. Lett. A 156 2

\bibitem {mosta1999}A. Mostafazadeh 1999, J. Phys. A 32 8325

\bibitem {glasser2009}M.L. Glasser, J. Mateo, J. Negro, and L.M. Nieto 2009,
Chaos Solitons Fract. 41 2067.

\bibitem {fojon2010}O. Fojon, M. Gadella and L. P. Lara 2010, Comput Math Appl
59, 264.

\bibitem {greenberger}D. M. Greenberger Physica B 151 (1988) 374

\bibitem {makowski1992}A. J. Makowski and P. Peplowski 1992, Phys. Lett. A 163 142.

\bibitem {zou2000}J. Zou and B. Shao 2000, Int. J. Mod. Phys. B 14 1059.

\bibitem {yao2001}Q.K. Yao Qian-Kai, G. W. Ma, X. F. Chen and Y. Yu 2001, Int.
J. Theor. Phys. 40 551.

\bibitem {wang2008}Z. S. Wang, C. Wu, X. L Feng , L.C. Kwek, C.H. Lai, C.H. Oh
and V. Vedral 2008, Phys. Lett. A 372 775

\bibitem {mousavi2012}S. V. Mousavi 2012, EPL 99 30002.

\bibitem {mousavi2014}S. V. Mousavi 2014, Phys. Scr. 89 065003.

\bibitem {facchi}P. Facchi, G. Garnero, G. Marmo and J. Samuel 2016, Ann.
Phys. 372 201

\bibitem {bellman}R. Bellman, \emph{A brief introduction to Theta functions}
(Dover: Mineola, NY, 2013).

\bibitem {aharonov-anandan}Y. Aharonov and J. Anandan 1987, Phys. Rev. Lett.
58 1593.

\bibitem {book-GP}A. Bohm, A. Mostafazadeh, H. Koizumi, Q. Niu and J.
Zwanziger, \emph{The Geometric Phase in Quantum Systems} (Springer, Berlin, 2003).

\bibitem {anandan}J. S. Anandan 1988, Ann. Inst. Henri Poincar\'{e} 49 271.

\bibitem {matzkin2012}A.\ Matzkin 2012, Phys.\ \ Rev.\ Lett. 109 150407

\bibitem {child98}Y. C. Ge and M. S. Child 1997, Phys. Rev. Lett. 78 2507.

\bibitem {A2015JPA}A.\ Matzkin 2015, J. Phys. A 48 305301
\end{thebibliography}
\end{document}